\documentclass[prd,superscriptaddress,twocolumn,superscriptaddress,amsmath]{revtex4-2}
\usepackage{graphicx}
\usepackage{float}
\usepackage{epstopdf,cancel}
\usepackage{epsf,latexsym,bbm,euscript}
\usepackage{amssymb,amsmath}
\usepackage{mathtools} 
\usepackage{times,graphics}
\usepackage{soul,xcolor}
\usepackage{mathtools}

\def\6{{\langle}}
\def\9{{\rangle}}
\newcommand{\defeq}{\vcentcolon=}
\newcommand{\eqdef}{=\vcentcolon}

\newcommand{\be}{\begin{equation}}
\newcommand{\ee}{\end{equation}}
\newcommand{\ba}{\begin{eqnarray}}
\newcommand{\ea}{\end{eqnarray}}

\def\etal{\textit{et al.}}

\usepackage{url,hyperref}
\hypersetup{colorlinks,linkcolor={blue!55!black},citecolor={red!45!black},urlcolor={blue!45!black},breaklinks=true}

\begin{document}

\title{Spin optics for gravitational waves}

\author{Pravin Kumar Dahal}
 \email{pravin-kumar.dahal@hdr.mq.edu.au}
\affiliation{School of Mathematical \& Physical Sciences, Macquarie University}

\date{\today}

\begin{abstract}
\textbf{Abstract:} We present the geometric optics expansion for circularly polarized gravitational waves on a curved spacetime background, to subleading order. We call spin optics to the subleading order geometric optics expansion, which involves modifying the standard eikonal function by including a specially chosen helicity-dependent correction. We show that the techniques developed for the propagation of electromagnetic waves can also be applied to gravitational waves in the limit of spin optics. However, one needs to account for the difference in the photon and graviton helicity, which we do here.
\end{abstract}

\keywords{Geometric optics; spin Hall effect; Gravitational Faraday rotation; Spin optics.}

\maketitle

\section{Introduction}

The geometric optics approximation can be applied to study high-frequency gravitational waves. It states that in the infinitely large frequency limit, gravitational waves in the fixed curved background behave as null ray trajectories, similar to electromagnetic waves~\cite[]{1,9,10}. The geometric optics expansion reduces the problem of solving the linearized Einstein field equations to solving the ray and transport equations along these rays. In the geometric optics limit, the backreaction onto the ray equations from the polarization vector is absent. Because of this, the laws of geometric optics are no longer valid at large but finite frequencies.

The backreaction from helicity might cause ray trajectories to deviate from geodesics appreciably. This effect, in which the propagation of gravitational waves in the subleading order in curved spacetime (orbital motion) depends on the helicity (spin), is also called the gravitational spin Hall effect~\cite[]{2,11,38,39}. Thus, the spin-orbit interaction modifies the gravitational wave propagation from the original ray trajectory as determined by geometric optics. Here, we present the covariant formulation of the Wentzel-Kramers-Brillouin (WKB) analysis for polarized gravitational waves in the subleading order geometric optics expansion in wavelength. We call the ``spin optics" approximation to the WKB analysis up to  subleading order in the geometric optics approximation. The procedure here will be similar to that developed for electromagnetic waves~\cite[]{11,14}.

The spin Hall effect results from the interaction of the polarization/spin with the orbital motion of the rays~\cite[]{13,15,16}. The effective ray equations describing this effect for gravitational waves are similar to those for light. The spin Hall effect for light (also called the optical Magnus effect~\cite[]{17}) is observed when electromagnetic waves travel in an inhomogeneous medium~\cite[]{18,19}. Spin-orbit coupling results from the interaction of the polarization degrees of freedom with the gradient of the refractive index of the medium. As a result, transverse deflection of electromagnetic waves in a direction perpendicular to the refractive index gradient occurs. The spin Hall effect for light can be explained in terms of Berry curvature~\cite[]{6,7} and provides a correction to the geometric optics approximation, which scales approximately as the inverse of the frequency (in the subleading order). One can reinterpret the spin Hall effect from condensed matter physics in the context of general relativity, where spacetime curvature itself plays the role of an inhomogeneous medium~\cite[]{2,5,14,20,21,22,23}. In this setting, the gravitational spin Hall effect arises from the interaction of polarization with the spacetime curvature itself, resulting in a spin-dependent correction of the particle dynamics.

A higher-order geometric optics expansion is not the only approach for calculating the spin Hall effect for gravitational waves. Effects of spin on the trajectory of massive particles have been worked out by \cite{29,30,31}. The Mathisson-Papapetrou-Dixon equations have been adapted to the massless case by the work of \cite{27,28}, and their equations can be used to study the polarization-dependent correction on the trajectory of massless particles (see \cite{5} and the reference therein for details).

This article is the study of the propagation of gravitational waves through spacetime curvature. Here we are not concerned about the two other subscopes of gravitational wave physics, its generation and detection. We also provide applications depicting how gravitational wave propagation evolves. Gravitational waves emitted by the source propagate over cosmological distances before reaching the observer. Such waves might encounter inhomogeneities in the form of spacetime curvature during their propagation. If the length scale of these inhomogeneities, which acts as a lensing object, is much larger than the characteristic wavelength of the gravitational waves, then the geometric approximation is valid. This is the standard approach to studying gravitational wave propagation. Geometric optics approximation implies that we treat gravitational wave as a particle propagating in the geodesic trajectory and its polarization tensor parallel transported along the geodesic. However, the subleading order correction from the geometric optics for gravitational waves might be necessary when studying gravitational lensing~\cite[]{24,25}, which necessitates calculating the higher-order geometric optics correction for gravitational waves. In realistic situations, there are gravitational wave detectors that can detect waves with a wavelength of astrophysical lengths. This implies that the waves under consideration will have wavelengths comparable to or even larger than the Schwarzschild radius of astrophysical objects, and the standard lensing phenomena will likely include wave effects that must be appropriately taken into account. Spin-orbit coupling plays a role when analyzing the propagation of spinning particles in an inhomogeneous medium when the wavelength of a particle is small but not negligible compared to the inhomogeneity scale of the medium.

This article is organized as follows: In Sec.~\ref{ee1} we linearize the Einstein field equations with some simplifying assumptions, giving rise to the equations governing the propagation of gravitational waves. Then in Sec.~\ref{so3}, we make a WKB approximation of the metric perturbation and substitute it back into linearized Einstein equations to obtain the gravitational wave solution in the high-frequency limit. In the leading order approximation in $1/\omega$, the results are consistent with the geometric optics approximation. This gravitational wave solution in the geometric optics limit can be represented by a set of null tetrads satisfying some orthonormal and completeness conditions. The scalar product of Fermi transported null tetrads satisfies those conditions everywhere along the null trajectory of gravitational waves, provided that they are satisfied initially. Next, we cast the gravitational wave equations in the form of Maxwell's equations for electromagnetic waves. In doing so, we can describe the circular polarization of the gravitational waves in terms of the self-dual and anti-self-dual solutions of some bivector/field tensor. After laying the foundations for the generalization of geometric optics, we will write the equations for the trajectory and polarization of gravitational waves in the subleading order approximation. In Sec.~\ref{a4}, we calculate the effect of spin-orbit coupling in a number of simple cases: 1) gravitational lensing of gravitational waves in Schwarzschild spacetime and 2) the propagation of gravitational waves in an expanding universe. Finally, we present some concluding remarks and discussions in Sec.~\ref{cd4}.

Throughout this article, we consider a Lorentzian manifold with the metric $g_{\mu\nu}$ of the signature $\left(-,+,+,+\right)$. Similarly, we write the complex conjugate of $z$ as $\tilde z$, adopt the Einstein summation convention and denote the covariant derivative by a semicolon ($;$) (except in Sec.~\ref{ee1} and the related appendix, where we use $\nabla$ for convenience). We use natural units where $G=c=1$. $\lambda$ denotes the affine parameter along gravitational wave trajectories and $\dot x= d x/d\lambda$. The sign convention used here is adopted by \cite{1}.

\section{Linearization of Einstein field equations}\label{ee1}

We begin by considering the Einstein field equations with vanishing cosmological constant
\begin{equation}
    R_{\alpha\beta}-\frac{1}{2}R g_{\alpha\beta}=0,
\end{equation}
where $R_{\alpha\beta}$ is the Ricci tensor, and $R$ denotes the Ricci scalar. Here, we describe the propagation of gravitational waves, treating it as a small metric perturbation around the fixed background solution of the vacuum Einstein equations. We assume that the fixed background, given by the metric $g_{\alpha\beta}$, describes the gravitational field of arbitrary strength and its perturbation is small, that is, $h_{\alpha\beta} \ll g_{\alpha\beta}$.

Let $g_{\alpha\beta}$ be the solution of Einstein equations in a vacuum
\begin{equation}
    R_{\alpha\beta}=0. \label{vs2}
\end{equation}
Consider another metric $\tilde g_{\alpha\beta}$, which is the result of a small perturbation $h_{\alpha\beta}$ on $g_{\alpha\beta}$
\begin{equation}
    \tilde g_{\alpha\beta}=g_{\alpha\beta}+h_{\alpha\beta}.
\end{equation}
Substituting this into Eq.~\eqref{vs2} gives, up to first order in $h_{\alpha\beta}$ and its derivatives (see \cite{2})
\begin{multline}
    \Big(\delta^\gamma_\alpha \delta^\delta_\beta \nabla_\mu \nabla^\mu- g_{\alpha\beta}g^{\gamma \delta}\nabla_\mu \nabla^\mu+ g^{\gamma\delta} \nabla_\alpha \nabla_\beta+ g_{\alpha\beta} \nabla^\gamma \nabla^\delta\\
    -\delta^\delta_\beta \nabla^\gamma \nabla_\alpha- \delta^\delta_\alpha \nabla^\gamma \nabla_\beta \Big)h_{\gamma\delta}=0.\label{fe4}
\end{multline}
Now, taking the trace of this equation yields
\begin{equation}
    \nabla^\alpha \nabla^\beta h_{\alpha\beta}-\nabla^\alpha \nabla_\alpha h^\mu_\mu=0.\label{tr5}
\end{equation}
Using the available gauge freedom, we can further reduce Einstein's field equations to a hyperbolic system. Let us choose the Lorentz gauge condition such that
\begin{equation}
    \nabla^\alpha h_{\alpha\beta}-\frac{1}{2}g_{\alpha\beta}\nabla^\alpha h^\mu_\mu=0.\label{lc6}
\end{equation}
Substituting this back into Eq.~\eqref{tr5} gives
\begin{equation}
    \nabla^\alpha \nabla_\alpha h^\mu_\mu =0.\label{tr7}
\end{equation}
Again substituting Eqs.~\eqref{lc6} and \eqref{tr7} into Einstein field Eq.~\eqref{fe4} gives (see Appendix~\ref{app0})
\begin{equation}
    \nabla^\mu \nabla_\mu h_{\alpha\beta}+ 2 R_{\alpha\mu\beta\nu}h^{\mu\nu}=0, \label{lfe8}
\end{equation}
where we have used the following relation expressing the commutation of covariant derivatives in terms of the Riemann curvature tensor $R_{\alpha\mu\beta\nu}$
\begin{equation}
    \nabla_\beta \nabla_\alpha h^\mu_\nu- \nabla_\alpha \nabla_\beta h^\mu_\nu= R^\sigma_{\nu\alpha\beta} h^{\mu}_\sigma- R^\mu_{\sigma\alpha\beta} h^{\sigma}_\nu. \label{covd9}
\end{equation}

\subsection{Initial conditions}\label{2i}

Here, we assume that the gradient of the trace of the perturbation tensor $h_{\alpha\beta}$ vanishes initially. Then, Eq.~\eqref{tr7} ensures that $\nabla^\alpha h^\mu_\mu$ vanishes everywhere along the trajectory. Thus, the field equations and the Lorentz gauge condition for gravitational waves reduce to
\begin{align}
    &\nabla^\mu \nabla_\mu h_{\alpha\beta}+ 2 R_{\alpha\mu\beta\nu}h^{\mu\nu}=0,\label{fe9}\\
    &\nabla^\alpha h_{\alpha\beta}=0,\label{lc10}
\end{align}
respectively. Note the striking similarity of these equations with Maxwell's equations and the Lorentz gauge condition for electromagnetic radiation. This will be the case for their solutions as well, as shown below. We consider circularly polarized gravitational waves for our calculations.

\section{Formulation of spin optics}\label{so3}

The spin optics approximation is valid when the typical wavelength of the waves is small (but cannot be neglected) compared to the length scale of the variation of its amplitudes and wavelength and the radius of curvature of the spacetime on which it propagates. In that limit, waves can be approximated locally as rays propagating on an approximately flat spacetime. The spin optics approximation can be expressed mathematically by using the WKB ansatz
\begin{equation}
    h^{\alpha\beta}=a^{\alpha\beta} e^{i \omega {\cal S}},\label{a11}
\end{equation}
where $a^{\alpha\beta}$ is the slowly varying complex amplitude, and $\omega {\cal S}$ is the rapidly varying real phase. Here, $\omega$ is the characteristic frequency of the problem. In later calculations, we write the wave vector $l_\alpha={\cal S}_{;\alpha}$, where semicolon $;$ denotes the covariant derivative. The wave vector and the polarization tensor can be expanded in terms of $\omega$ as
\begin{align}
    & l^\alpha=l^\alpha_0+\frac{l^\alpha_1}{\omega}+\frac{l^\alpha_2}{\omega^2}+ ...,\\
    & a^{\alpha\beta}=a^{\alpha\beta}_0+\frac{a^{\alpha\beta}_1}{\omega}+\frac{a^{\alpha\beta}_2}{\omega^2}+ ...,
\end{align}
Below in Sec.~\ref{ntb}, we will use the Fermi propagated null tetrad. Two of its components represent solutions for the trajectory and polarization of gravitational waves. This Fermi propagation reduces the freedom in the transformation property of the null tetrad $m^\alpha \to e^{i {\cal S}_1 (\lambda)} m^\alpha$ by the condition
\begin{equation}
    \frac{d {\cal S}_1 (\lambda)}{d\lambda}= 0
\end{equation}
This is why we cannot absorb higher-order phase factors such as ${\cal S}_1(\lambda)$ into the complex amplitude $a^{\alpha\beta}_0$ by such transformations. We can recover the equations of geometric optics by substituting this perturbation metric onto the source free wave equation (Eq.~\eqref{fe9}) and the Lorentz gauge condition. Let us first start from the Lorentz condition, which can be written as
\begin{equation}
    l_0^\alpha a_{0 \alpha\beta}+\frac{1}{\omega}\left(l_0^\alpha a_{1 \alpha\beta}+l_1^\alpha a_{0 \alpha\beta}- i a^\alpha_{0~\beta;\alpha}\right)=0,\label{lc14}
\end{equation}
up to subleading order in $\omega$. We next substitute the perturbation tensor into the source-free wave equation, which again, up to subleading order in $\omega$ gives
\begin{multline}
    t^{\alpha\beta} \eqdef a_0^{\alpha\beta} l_{0\mu} l_0^\mu+\frac{1}{\omega}\bigg(a_1^{\alpha\beta} l_{0\mu} l_0^\mu+ 2 a_0^{\alpha\beta} l_{1\mu} l_0^\mu\\
    -i \left(a_0^{\alpha\beta} l^\mu_{0;\mu}+2 a^{\alpha\beta}_{0~~;\mu} l_0^\mu \right) \bigg)=0.\label{fe15}
\end{multline}
Let us now calculate the identically vanishing quantity $\tilde a_{0\alpha\beta}t^{\alpha\beta}+ a_{0\alpha\beta} \tilde t^{\alpha\beta}$, which gives the dispersion relation
\begin{equation}
     l_{0\beta} l_0^\beta+\frac{2}{\omega} \left(l_{1 \beta}-b_\beta \right) l_0^\beta =0.\label{dr16}
\end{equation}
In the above expression, we have used $\tilde a_0^{\alpha\beta} a_{0 \alpha\beta}=a^2$ and substituted
\begin{equation}
    \frac{i}{2 a^2}\left(\tilde a^{\alpha\beta} a_{\alpha\beta;\mu}- a^{\alpha\beta} \tilde a_{\alpha\beta;\mu} \right) \defeq b_\mu.\label{b17}
\end{equation}

\subsection{Introduction of null tetrad}\label{ntb}

In the geometric optics limit, the wave vector $l_0^\alpha$ and the polarization vector $m_0^\alpha$ satisfying the wave equations can be identified with two components of a null tetrad (see Appendix~\ref{a31}). The remaining two components would then be a nonunique auxiliary null vector $n_0^\alpha$ and a complex conjugate $\tilde m_0^\alpha$. A set of null tetrads $e^\alpha_i= \left(l_0^\alpha, n_0^\alpha, m_0^\alpha, \tilde m_0^\alpha\right)$, where $i=\{l_0,n_0,m_0,\tilde m_0\}$ represents the tetrad indices, would satisfy the following orthogonality and completeness relationships
\begin{align}
    & l_0^\alpha m_{0 \alpha}= l_0^\alpha l_{0\alpha}= l_0^\alpha \tilde m_{0\alpha}=0,  \qquad m_0^\alpha \tilde m_{0 \alpha}=1, \label{oc29}\\
    & m_0^\alpha m_{0 \alpha}= \tilde m_0^\alpha \tilde m_{0\alpha}=0,\label{oc30}\\
    & n_0^\alpha m_{0 \alpha}= n_0^\alpha n_{0\alpha}= n_0^\alpha \tilde m_{0\alpha}=0, \qquad n_0^\alpha l_{0\alpha}=-1. \label{oc31}
\end{align}
Here Eqs.~\eqref{oc29} follows from the geometric optics approximation (see Eqs.~\eqref{go18}) and Eqs.~\eqref{oc30} and \eqref{oc31} are purely by choice of $n_{0 \alpha}$ and $m_{0 \alpha}$. In Sec.~\ref{pb}, we will show that choosing $m_{0 \alpha}$ to satisfy Eqs.~\eqref{oc30} is equivalent to choosing the polarization state. In particular, these relations hold for circularly polarized waves. Moreover, these components of the null tetrad also satisfy
\begin{align}
    & l^\alpha_{0;\beta}l_0^\beta=0, \qquad m^\alpha_{0;\beta}l_0^\beta=0=\tilde m^\alpha_{0;\beta}l_0^\beta, \label{pt32}\\
    & n^\alpha_{0;\beta}l_0^\beta=0 ,\label{pt33}
\end{align}
where Eqs.~\eqref{pt32} are again from the geometric optics approximation (see Eq.~\eqref{go28}), and $l_{0 \alpha;\beta}=l_{0 \beta;\alpha}$ is used in obtaining the first relation. Eq.~\eqref{pt33} is obtained by choice of $n_0^\alpha$. To show that this choice is indeed possible, we introduce the Fermi derivative operator ${\cal D}_l$ along the ray $l^\alpha$, which gives the following relation when applied to the tensor $A^\alpha$~\cite[]{14}
\begin{equation}
    {\cal D}_l A^\alpha =l_0^\beta A^\alpha_{;\beta}-w_\beta A^\beta n^\alpha+ A^ \beta n_{ \beta} w^\alpha,
\end{equation}
where $w^\alpha =l_0^\beta l^\alpha_{;\beta}$. $w^\alpha$ is an identically vanishing quantity in geometric optics. As $l^\alpha l_\alpha=0$, we have ${\cal D}_l l^\alpha=0$. A vector $A^\alpha$ is Fermi propagated if its Fermi derivative ${\cal D}_l A^\alpha=0$, and it is easy to see that the scalar product of any two Fermi propagated vectors is constant. From this, we can conclude that if the tetrad $\left(l^\alpha, n^\alpha, m^\alpha, \tilde m^\alpha\right)$ satisfies the orthogonality and completeness relations similar to the one given in Eqs.~\eqref{oc29}-\eqref{oc31} at some point on the ray, and if they are Fermi propagated, they satisfy those relations everywhere on the ray. Therefore, we Fermi propagate the frame of a null tetrad $\left(l^\alpha, n^\alpha, m^\alpha, \tilde m^\alpha\right)$ so that they satisfy the orthogonality and completeness relations analogous to those in Eqs.~\eqref{oc29}-\eqref{oc31} along the ray. The components of this tetrad also obey
\begin{align}
    & l_0^\beta n^\alpha_{;\beta}= w^\beta n_{ \beta} n^\alpha,\\
    & l_0^\beta m^\alpha_{;\beta}= w^\beta m_{ \beta} n^\alpha,\\
    & l_0^\beta \tilde m^\alpha_{;\beta}= w^\beta \tilde m_{ \beta} n^\alpha.
\end{align}
We can now use the freedom in the choice of null tetrad
\begin{equation}
    l^\alpha \to A l^\alpha, n^\alpha \to A^{-1} n^\alpha,
\end{equation}
to fix $w^\beta n_{ \beta}=0$, where $A$ is a real function. This condition fixes the parameter $\lambda$ along the ray up to its possible rescaling $\lambda \to A^{-1}\lambda$, and we call such a choice a canonical parametrization~\cite[]{14}. Therefore, in the canonical parametrization, we have
\begin{equation}
    l_0^\beta n^\alpha_{;\beta}= 0, \qquad
    l_0^\beta m^\alpha_{;\beta}= w^\beta m_{ \beta} n^\alpha,\qquad
    l_0^\beta \tilde m^\alpha_{;\beta}= w^\beta \tilde m_{ \beta} n^\alpha.\label{pe39}
\end{equation}
From these results, we can see that a null tetrad $\left(l_0^\alpha, n_0^\alpha, m_0^\alpha, \tilde m_0^\alpha\right)$ satisfying evolution Eqs.~\eqref{pt32}-\eqref{pt33} in the geometric optics limit naturally generalizes to a null tetrad $\left(l^\alpha, n^\alpha, m^\alpha, \tilde m^\alpha\right)$ satisfying evolution Eqs.~\eqref{pe39}.

\subsection{Introduction of polarization basis}\label{pb}

In the geometric optics limit, circularly polarized gravitational waves have amplitude $a_{0 \alpha\beta}= a_0 m_{0\alpha} m_{0\beta}$ whose evolution is governed by the equation $m^\alpha_{0;\beta}l_0^\beta=0$ (see Appendix~\ref{a31}). This means that the polarization is parallel propagated along the trajectory in geometric optics. The same line of arguments used from Eqs.~\eqref{go18}-\eqref{pa23} can be used to show that, in the spin optics approximation, the polarization tensor can be written in the form
\begin{equation}
    a_{\alpha\beta}= a m_{\alpha} m_{\beta},
\end{equation}
provided that an arbitrary vector field $\xi^\alpha$ in Eq.~\eqref{b3} satisfies
\begin{equation}
    \xi_\alpha l^\alpha l_\beta-\frac{i}{\omega}  \left(\xi_{\alpha;\beta}+\xi_{\beta;\alpha}\right) l^\alpha - \frac{i}{\omega} \left(\xi_\beta l^\alpha+ \xi^\alpha l_\beta \right)_{;\alpha}= 0.
\end{equation}
We can keep track of the evolution of this polarization tensor by defining an antisymmetric field analogous to electromagnetism,
\begin{equation}
    F_{\beta\gamma} \defeq A_{\gamma;\beta}- A_{\beta;\gamma},\label{ft40}
\end{equation}
where the vector $A_\gamma$ is given as
\begin{equation}
    A_\gamma \defeq a m_\gamma e^{i {\cal S}/2}.\label{vp41}
\end{equation}
The geometric optics approximation up to subleading order in $1/\omega$ assumes that the typical length scale of variation of amplitude, polarization, and wavelength of gravitational waves is negligible compared to the radius of curvature of the spacetime through which the waves propagate. In this approximation, the curvature terms appearing in the linearized gravitational wave equation Eq.~\eqref{fe9} can be safely neglected, as seen from Eq.~\eqref{fe15}. In such a regime, the antisymmetric field $F_{\beta\gamma}$ satisfies
\begin{align}
    &F_{\alpha\beta;\gamma}+F_{\gamma\alpha;\beta}+F_{\beta\gamma;\alpha}=0, \label{fe41}\\
    &F^{\beta\gamma}_{~~~~;\gamma}=0.\label{fe42}
\end{align}
Let us define the complex version of the tensor $F^{\alpha\beta}$ as
\begin{equation}
    {\cal F}^\pm=F\pm i F^*,
\end{equation}
where $F^*=\epsilon_{\alpha\beta\mu\nu} F^{\mu\nu} /2$ is the Hodge dual of $F^{\alpha\beta}$. Here $\epsilon_{\alpha\beta\mu\nu}$ is the Levi-Civita tensor in four dimensions, and its components in the tetrad basis are $i l \wedge n \wedge m \wedge \tilde m$.
As $(F^*)^*=-F$, we have $\left({\cal F}^\pm\right)^*= \mp i {\cal F}^\pm$, a property by virtue of which we call ${\cal F}^{\pm}$ self-/anti-self-dual antisymmetric field for $+/-$ sign. Now, the field ${\cal F}^{+}_{\alpha\beta}$ can be expanded by substituting Eq.~\eqref{vp41} into Eq.~\eqref{ft40}
\begin{equation}
    {\cal F}^{+}_{\alpha\beta}= \frac{i \omega}{2} {\cal Z}_{\alpha\beta} e^{i {\cal S}/2},
\end{equation}
where
\begin{multline}
    {\cal Z}_{\alpha\beta}=a \bigg(l_\beta m_\gamma-l_\gamma m_\beta -\frac{2 i}{\omega}\Big(\frac{a_{;\beta} }{a} m_\gamma -\frac{a_{;\gamma}}{a} m_\beta\\
    + m_{\gamma;\beta}- m_{\beta;\gamma}\Big)\bigg). \label{fa45}
\end{multline}
For this self-dual field ${\cal F}^{+}_{\alpha\beta}$, using the property that contraction of a self-dual field with an anti-self-dual field vanishes, we obtain
\begin{align}
    &{\cal Z}_{\alpha\beta} m^\alpha n^\beta=0,\label{pe46}\\
    &{\cal Z}_{\alpha\beta} \left(\tilde m^\alpha m^\beta- l^\alpha n^\beta \right)=0,\label{pe47}\\
    &{\cal Z}_{\alpha\beta} l^\alpha \tilde m^\beta=0, \label{pe48}
\end{align}
By substituting the value of ${\cal Z}_{\alpha\beta}$ from Eq.~\eqref{fa45}, we can see that Eq.~\eqref{pe48} is identically satisfied in the limit of geometric optics. However, Eqs.~\eqref{pe46} and \eqref{pe47} gives
\begin{equation}
    m_0^\alpha m_{0 \alpha}=0 = l_0^\alpha m_{0\alpha}
\end{equation}
These are the orthogonality relations given in Eqs.~\eqref{oc29} and \eqref{oc30}.

\subsection{Equations of spin optics}\label{soD}

\subsubsection{Defining the Hamiltonian}

We have shown in Sec.~\ref{ntb} that in the geometric optics limit, light travels along the null geodesic
\begin{equation}
    l_0^\alpha l_{0\alpha}=0, \qquad l^\alpha_{0;\beta}l_0^\beta=0.\label{go51}
\end{equation}
Let us take $x^\alpha(\lambda)$ to be the integral curve of $l^\alpha$. Then
\begin{equation}
    l_0^\alpha= \frac{d x_0^\alpha}{d \lambda} \equiv \dot x_0^\alpha.
\end{equation}
Eq.~\eqref{go51} can be recovered from Hamilton's equations of motion if we define the Hamiltonian as
\begin{equation}
    H= \frac{1}{2}g^{\alpha\beta} l_{0\alpha} l_{0 \beta},\label{ham53}
\end{equation}
Hamilton's equations of motion are
\begin{equation}
    \frac{d x_0^\alpha}{d \lambda}= \frac{\partial H}{\partial l_{0\alpha}}= g^{\alpha\beta} l_{0 \beta}, \label{he54}
\end{equation}
and
\begin{equation}
    \frac{d l_{0\alpha}}{d \lambda}= \frac{\partial H}{\partial x_0^\alpha}=  \frac{1}{2} \dot x_0^\mu \dot x_0^\nu \frac{\partial g_{\mu\nu}}{\partial x_0^\alpha},
\end{equation}
where we have used Eq.~\eqref{he54} and the relation
\begin{equation}
    \frac{\partial g_{\alpha\beta}}{\partial x_{0\mu}}= -g_{\nu\alpha} g_{\rho\beta}\frac{\partial g^{\nu\rho}}{\partial x_{0\mu}}.
\end{equation}
in obtaining this. Further simplification of this equation gives
\begin{equation}
    \frac{d \left(g_{\alpha\beta} \dot x_0^\beta\right)}{d \lambda}- \frac{1}{2} \dot x_0^\mu \dot x_0^\nu \frac{\partial g_{\mu\nu}}{\partial x_{0\alpha}}= \frac{D^2 x_0^\alpha}{D\lambda^2}=0,\label{he57}
\end{equation}
where $D/D\lambda$ denotes the covariant derivative along the curve $x^\alpha(\lambda)$. The two Eqs.~\eqref{he54} and \eqref{he57} are identical to Eq.~\eqref{go51}, thereby demonstrating that the Hamiltonian defined in Eq.~\eqref{ham53} correctly reproduces the geometric optics equations. The action corresponding to this Hamiltonian is
\begin{equation}
    {\cal A}_0= \frac{1}{2} \int \dot x_0^\alpha \dot x_{0\alpha} d\lambda.
\end{equation}
This action corresponds to the dispersion relation in the limit of geometric optics, and this dispersion relation is Eq.~\eqref{dr16} in the leading order approximation. Thus, because of the subleading order terms of the dispersion relation Eq.~\eqref{dr16}, the natural generalization to spin optics would give an action of the form
\begin{equation}
    {\cal A}= \frac{1}{2} \int \dot x^\alpha \dot x_\alpha d\lambda + \frac{1}{\omega} \int b_\alpha \dot x^\alpha d\lambda.
\end{equation}
This action is analogous to that of electromagnetic waves considered in \cite{37} and \cite{14}, where the spin Hall effect of light is derived. The variation of the first term, which is the optical path length, gives
\begin{equation}
    \frac{1}{2} \delta\int \dot x^\alpha \dot x_\alpha d\lambda= \int \dot x_\alpha \frac{D \delta x^\alpha}{D\lambda} d\lambda= - \int \frac{D^2 x_ \alpha}{D\lambda^2} \delta x^\alpha d\lambda.
\end{equation}
Similarly the variation of the second term (which resembles the Berry connection in optics) gives
\begin{multline}
    \frac{1}{\omega} \delta\int b_\alpha \dot x^\alpha d\lambda= \frac{1}{\omega} \int \delta b_\alpha \dot x^\alpha d\lambda+ \frac{1}{\omega} \int b_\alpha \frac{D \delta x^\alpha}{D\lambda} d\lambda\\
    = \frac{1}{\omega} \int b_{\alpha;\beta} \dot x^\alpha \delta x^\beta d\lambda- \frac{1}{\omega} \int b_{\beta;\alpha} \delta x^\beta \dot x^\alpha d\lambda.
\end{multline}
The application of the variational principle $\delta{\cal A}=0$ yields
\begin{equation}
    \frac{D^2 x_\beta}{D\lambda^2}+ \frac{1}{\omega} \left( b_{\beta;\alpha}- b_{\alpha;\beta}\right) \dot x^\alpha= 0. \label{vp62}
\end{equation}
We can write the corresponding Hamiltonian as
\begin{equation}
    H= \dot x^\alpha l_\alpha- {\cal L},\label{h63}
\end{equation}
where ${\cal L}= \dot x^\alpha \dot x_\alpha/2 + b_\alpha \dot x^\alpha/ \omega$ is the Lagrangian. We substitute the canonical momentum
\begin{equation}
    l_\alpha= \frac{\partial {\cal L}}{ \partial \dot x^\alpha}= \dot x_\alpha + \frac{b_\alpha}{\omega}, \label{tr64}
\end{equation}
into Eq.~\eqref{h63} to get the Hamiltonian
\begin{equation}
    H= \frac{1}{2\omega^2}\left(\omega l_{0 \alpha} + l_{1\alpha}- b_\alpha\right) \left(\omega l_0^\alpha + l_1^\alpha- b^\alpha\right).\label{he65}
\end{equation}

\subsubsection{Solving Hamilton's equations of motion}

Now, we substitute Eq.~\eqref{tr64} into the dispersion relation of Eq.~\eqref{dr16} to obtain
\begin{equation}
    \left(l_{0\beta}+\frac{1}{\omega}(l_{1\beta}-b_\beta) \right) \left(l_0^\beta +\frac{1}{\omega}(l_1^\beta-b^\beta) \right)=\dot x_\beta \dot x^\beta=0.\label{dr50}
\end{equation}
Thus, in the leading order approximation in $1/\omega$, $\dot x^\alpha= l_0^\alpha$ gives the tangent vector. This equation implies that the gravitational wave trajectory in the spin optics approximation is still null. It is not geodesic however, as seen from the fact that
\begin{equation}
    \frac{D\dot x^\alpha}{D\lambda}=\frac{1}{\omega} \left(l^\alpha_{1;\beta}-b^\alpha_{;\beta}\right)l_0^\beta.\label{gd51}
\end{equation}
We use the Hamiltonian of Eq.~\eqref{he65} to calculate this quantity. This Hamiltonian has been solved by \cite{14} to obtain the following equation for right-hand circularly polarized rays:
\begin{equation}
    \frac{D^2 x^\alpha}{D\lambda^2}=\frac{1}{\omega} k^\alpha_\beta l_0^\beta,
\end{equation}
where $k_{\alpha \beta}=b_{\beta;\alpha}-b_{\alpha;\beta}$ and $b_\alpha= 2 i \tilde m_0^\beta m_{0 \beta;\alpha}$. This is our Eq.~\eqref{vp62}, which has been derived by using the variational principle. Comparing this with Eq.~\eqref{gd51}, we get $l_{1 \alpha;\beta}=b_{\beta;\alpha}$. Further simplifying $k_{\alpha\beta}$, one gets
\begin{equation}
    k_{\alpha\beta}=-2 i R_{\alpha\beta\mu\nu}m^\mu \tilde m^\nu+ 2 i \left(\tilde m^\mu_{;\alpha} m_{\mu;\beta}- \tilde m^\mu_{;\beta} m_{\mu;\alpha}\right).
\end{equation}
Substituting this back into Eq.~\eqref{gd51} gives
\begin{equation}
    \frac{D^2 x^\alpha}{D\lambda^2}=-\frac{2 i}{\omega} R^\alpha_{~\beta\mu\nu}m^\mu \tilde m^\nu l_0^\beta \approx -\frac{2 i}{\omega} R^\alpha_{~\beta\mu\nu} l_0^\beta m_0^\mu \tilde m_0^\nu.\label{te56}
\end{equation}
We thus obtain a nongeodesic trajectory in the spin optics approximation.

\subsubsection{Polarization equation}

We can substitute $w^\alpha =l_0^\beta l^\alpha_{;\beta}$ from Eq.~\eqref{gd51} into the Eqs.~\eqref{pe39}, which gives the evolution of the polarization vector
\begin{align}
    l_0^\beta n^\mu_{;\beta}= &0,\label{pe56}\\
    l_0^\beta m^\mu_{;\beta}= &\frac{1}{\omega} \left(l^\beta_{1;\alpha}- b^\beta_{;\alpha}\right)l_0^\alpha m_{ \beta} n^\mu\nonumber\\
    =&\frac{2 i}{\omega} R_{\alpha\beta\gamma\delta} l_0^\alpha m_0^\beta m_0^\gamma \tilde m_0^\delta n_0^\mu,\label{pe57}\\
    l_0^\beta \tilde m^\mu_{;\beta}= &\frac{1}{\omega} \left(l^\beta_{1;\alpha}- b^\beta_{;\alpha}\right)l_0^\alpha \tilde m_{ \beta} n^\mu \nonumber\\
    =&-\frac{2 i}{\omega} R_{\alpha\beta\gamma\delta} l_0^\alpha \tilde m_0^\beta \tilde m_0^\gamma m_0^\delta n_0^\mu.\label{pe58}
\end{align}
These equations ensure that the set of tetrads $\left(\dot x^\alpha, n^\alpha, m^\alpha, \tilde m^\alpha\right)$ satisfy the normalization and orthogonality relations in Eqs.~\eqref{oc29}-\eqref{oc31} throughout the ray up to subleading order in $1/\omega$. To see this, let us first simplify Eq.~\eqref{pe57} as
\begin{equation}
    l_0^\beta m^\mu_{;\beta} \approx \frac{1}{\omega} l_0^\beta m^\mu_{1;\beta}\approx \frac{1}{\omega} l_0^\beta \left(l_1^\alpha- b^\alpha\right)_{;\beta} m_{0\alpha} n_0^\mu.
\end{equation}
As the covariant derivatives of $m_{0\alpha}$ and $n_0^\mu$ are zero, we can write
\begin{equation}
    m_1^\mu= \left(l_1^\alpha-b^\alpha \right)m_{0 \alpha} n_0^\mu.\label{so60}
\end{equation}
Simplification of Eqs.~\eqref{pe56} and \eqref{pe58} in a similar manner gives
\begin{equation}
    n_1^\mu=0, \qquad \tilde m_1^\mu= \left(l_1^\alpha-b^\alpha \right) \tilde m_{0 \alpha} n_0^\mu,\label{so61}
\end{equation}
respectively. One can also write the subleading order of the component of tetrad $\dot x^\alpha$ differently by noting that $w^\alpha l_{0\alpha}=0=w^\alpha n_{0\alpha}$. This is how we have constructed the Fermi propagated tetrad in Sec.~\ref{ntb}. Therefore, $w^\alpha$ can be written in the form
\begin{align}
    &w^\alpha \equiv \frac{1}{\omega} l_0^\beta \left(l_1^\alpha- b^\alpha\right)_{;\beta}=-\tilde \kappa m_0^\alpha- \kappa \tilde m_0^\alpha,\\ &\kappa= - \frac{1}{\omega} m_0^\alpha l_0^\beta \left(l_{1\alpha}- b_\alpha\right)_{;\beta}.\nonumber
\end{align}
From this, one can write
\begin{equation}
    l_1^\alpha- b^\alpha= \tilde m_0^\beta \left(l_{1\beta}- b_\beta\right) m_0^\alpha + m_0^\beta \left(l_{1\beta}- b_\beta\right) \tilde m_0^\alpha. \label{so63}
\end{equation}
It is easy to see that Eq.~\eqref{so60}, in conjunction with the Eqs.~\eqref{so61} and \eqref{so63} taken as the subleading order correction of the tetrad $\left(\dot x^\alpha, n^\alpha, m^\alpha, \tilde m^\alpha\right)$ satisfies the scalar products of Eqs.~\eqref{oc29}-\eqref{oc31}. The leading order terms of this tetrad are $\left(l_0^\alpha, n_0^\alpha, m_0^\alpha, \tilde m_0^\alpha\right)$ of Sec.~\ref{ntb}. Moreover, the subleading order terms of the Lorentz gauge condition of Eq.~\eqref{lc14} gives
\begin{multline}
    \frac{a_{;\alpha}}{a}m_0^\alpha=- m^\alpha_{0; \alpha}- i b_\alpha m_0^\alpha+ m_0^\alpha \tilde m_{0\beta;\alpha} m_0^\beta\\
    = - m^\alpha_{0; \alpha}- \frac{i}{2} b_\alpha m_0^\alpha .\label{lc64}
\end{multline}

All of the polarization Eqs.~\eqref{pe46}-\eqref{pe48} are not satisfied in the limit of spin optics, thereby implying that the field, with the subleading order solution given by Eq.~\eqref{so60}, Eqs.~\eqref{so61} and \eqref{so63}, is not self-dual (see Appendix~\ref{aa}). However, a self-dual solution of these field equations should exist in the subleading order geometric optics approximation. We can find this self-dual solution by first writing the Fermi-like derivative operator as follows:
\begin{multline}
    {\cal D}'_l A^\alpha =l_0^\beta A^\alpha_{;\beta}-w_\beta A^\beta n^\alpha+ A^ \beta n_{ \beta} w^\alpha\\
    -\frac{2 i} {\omega} \left(\lambda_{;\mu}l^\mu m_\beta A^\beta m^\alpha- \tilde \lambda_{;\mu}l^\mu \tilde m_\beta A^\beta \tilde m^\alpha\right),
\end{multline}
where $\tilde \lambda= m_0^\alpha n_{0\alpha;\beta} m_0^\beta$. The vanishing of this derivative ${\cal D}'_l A^\alpha=0$ implies
\begin{multline}
    l_0^\beta A^\alpha_{;\beta}= w_\beta A^\beta n^\alpha- A^ \beta n_{ \beta} w^\alpha\\
    +\frac{2 i} {\omega} \left(\lambda_{;\mu}l^\mu m_\beta A^\beta m^\alpha- \tilde \lambda_{;\mu}l^\mu \tilde m_\beta A^\beta \tilde m^\alpha\right),
\end{multline}
and in that case, the scalar product of any two components of a tetrad $\left(\dot x^\alpha, n^\alpha, m^\alpha, \tilde m^\alpha\right)$ are constant except that of $m^\alpha$ with itself and that of $\tilde m^\alpha$ with itself. This can be seen by calculating
\begin{multline}
    \left(u^\alpha v_\alpha\right)_{;\beta}l_0^\beta=u^\alpha v_{\alpha_;\beta}l_0^\beta+ v^\alpha u_{\alpha_;\beta}l_0^\beta\\
    = \frac{4 i} {\omega}\left( \lambda_{;\mu}l^\mu m_\beta u^\beta m^\alpha v_\alpha- \tilde \lambda_{;\mu}l^\mu \tilde m_\beta u^\beta \tilde m^\alpha v_\alpha\right).
\end{multline}
This scalar product is non-zero only if $u^\alpha =v^\alpha=m^\alpha$ or $u^\alpha =v^\alpha=\tilde m^\alpha$. A tetrad with a vanishing Fermi-like derivative satisfies these conditions as well as the following orthogonality and completeness relations everywhere on the ray if they satisfy them at some point on the ray:
\begin{align}
    & \dot x^\alpha m_{\alpha}= \dot x^\alpha \dot x_\alpha= \dot x^\alpha \tilde m_{\alpha}=0,  \qquad m^\alpha \tilde m_{ \alpha} =1,\\
    & n^\alpha m_{\alpha}= n^\alpha n_{\alpha}= n^\alpha \tilde m_{\alpha}=0, \qquad n^\alpha l_{\alpha}=-1,
\end{align}
However, in general, $m^\alpha m_{\alpha}\ne 0$ and $\tilde m^\alpha \tilde m_{\alpha}\ne 0$  along the circularly polarized ray in the subleading order approximation. That is, the polarization vectors are not null like they are in the geometric optics approximation. Moreover, the tetrad evolves as
\begin{align}
    & l_0^\beta n^\alpha_{;\beta}= 0,\\
    & l_0^\beta m^\alpha_{;\beta}= w^\beta m_{ \beta} n^\alpha- \frac{2 i}{\omega} \tilde \lambda_{;\mu}l^\mu \tilde m^\alpha ,\label{pe74}\\
    & l_0^\beta \tilde m^\alpha_{;\beta}= w^\beta \tilde m_{ \beta} n^\alpha+ \frac{2 i}{\omega} \lambda_{;\mu}l^\mu m^\alpha.
\end{align}
The following simplification procedure is used to obtain Eqs.~\eqref{so60} and \eqref{so61} gives
\begin{align}
    &m_1^\mu= \left(l_1^\alpha-b^\alpha \right)m_{0 \alpha} n_0^\mu- 2 i \tilde \lambda \tilde m_0^\mu, \qquad n_1^\mu=0, \label{pe64}\\
    &\tilde m_1^\mu= \left(l_1^\alpha-b^\alpha \right) \tilde m_{0 \alpha} n_0^\mu+ 2 i \lambda m_0^\mu.
\end{align}
These components of tetrads constitute a solution of the linearized Einstein field equations in Lorentz gauge and are right-hand circularly polarized, as they also satisfy Eqs.~\eqref{pe46}-\eqref{pe48}. Therefore, they are the solutions for the propagation of circularly polarized gravitational waves in curved spacetime in the spin optics approximation.

\section{Applications}\label{a4}

\subsection{Gravitational lensing of gravitational waves}

As a first application of the spin-optics, we consider a localized object acting as a lens for gravitational waves of frequency $\omega$. The point-like lensing object is described by a Schwarzschild geometry of mass $M$. The geometric optics expansion for gravitational waves is valid only when their wavelength is much smaller than the Schwarzschild radius of the lensing body ($1/\omega \ll M$)~\cite[]{32,33}. Detectors like pulsar timing arrays (PTAs)~\cite[]{34} and the Laser Interferometer Space Antenna (LISA)~\cite[]{35} can collectively detect gravitational waves across a wide range of wavelengths, ranging from $10^8 m$ to $10^{17} m$ (the corresponding frequency range of $10^{-9}-1 Hz$). Thus, the lensing of real gravitational waves by the astrophysical objects in the wide mass range, including the galaxies, is likely to contain effects not seen in the geometric optics approximation~\cite{24}.

Here, we calculate the effect of spin-orbital coupling on gravitational wave lensing. In the geometric optics regime, the angle of deflection of gravitational waves by gravitating objects is the same as that of electromagnetic waves, and the maximum angle of deflection is given by
\begin{equation}
    \Delta \phi \approx \frac{4 M}{R_0},
\end{equation}
where $M$ is the mass and $R_0$ is the radius of a gravitating object.

For the Schwarzschild geometry, the general vector tangent to the congruence of null geodesics can be written as
\begin{equation}
    l_{0 \mu}=\left(-E,\frac{1}{1-\frac{2 M}{r}}\sqrt{E^2-\frac{L^2}{r^2}\left(1-\frac{2 M}{r}\right)},0,L \sin \theta \right),\label{tv72}
\end{equation}
where $E$ and $L$ are the constants of motion whose values are determined by the initial conditions of the geodesic trajectory. The three other null fields $n^\mu$, $m^\mu$ and $\tilde m^\mu$, along with $l^\mu$, form a set of null tetrads and satisfy the orthogonality and completeness relations given as
\begin{widetext}
\begin{equation}
    \begin{split}
        & n_{0 \mu}=\frac{1-\frac{2 M}{r}}{2\left(E^2-\frac{L^2}{r^2}\left(1-\frac{2 M}{r}\right)\right)}\left(-E,-\frac{1}{1-\frac{2 M}{r}}\sqrt{E^2-\frac{L^2}{r^2}\left(1-\frac{2 M}{r}\right)},0,L \sin \theta \right),\\
        & m_{0 \mu}=\frac{r}{\sqrt{2}} \left(-\frac{i L \left(1-\frac{2 M}{r}\right)}{r^2 \sqrt{E^2-\frac{L^2}{r^2}\left(1-\frac{2 M}{r}\right)}},0,1,\frac{i E \sin \theta} {\sqrt{E^2-\frac{L^2}{r^2}\left(1-\frac{2 M}{r}\right)}}\right),\qquad \tilde m_{0 \mu}= \left(m_{0 \mu}\right)^*. \label{t73}
    \end{split}
\end{equation}
\end{widetext}
We now apply the following transformation relations
\begin{align}
    & l_0^\mu\to l_0^\mu, \qquad m_0^\mu\to m_0^\mu+a l_0^\mu,\nonumber\\
    &n_0^\mu\to n_0^\mu+\tilde a m_0^\mu+ a \tilde m_0^\mu+ a \tilde a l_0^\mu,
\end{align}
to the null tetrad such that the resulting tetrad will be Fermi propagated along the rays. Here, the value of $a$ can be evaluated in such a way that the transformed vector $m_0^\mu+a l_0^\mu$ has vanishing covariant derivative to leading order:
\begin{equation}
    \left(m_0^\mu+a l_0^\mu\right)_{;\nu}l_0^\nu=m^\mu_{0;\nu}l_0^\nu+a_{,\nu}l_0^\mu l_0^\nu=0.
\end{equation}
In a spherically symmetric spacetime, $a$ can only be a function of $r$, thereby giving
\begin{multline}
    a=\int{\frac{m^\mu_{0;\nu}l_0^\nu n_{0 \mu}}{l_0^r}}dr\\
    =\int{\frac{i E L (r-3 M)}{\sqrt{2} r^3 \left(E^2-\frac{L^2}{r^2} \left(1-\frac{2 M}{r}\right)\right)^{3/2}}}dr.
\end{multline}
These transformed null tetrads satisfy Eqs.~\eqref{oc29}-\eqref{pt33}. These null trajectories $l_0^\mu$ and polarization vectors $m_0^\mu$ constitute solutions of the linearized Einstein equations in the geometric optics limit.

Now, to obtain the first-order correction to the null trajectory, we substitute these values into the propagation Eqs.~\eqref{te56}, and we get
\begin{align}
    &\frac{D^2 x^2}{D\lambda^2}=\frac{6 i L M}{\omega}\left(\frac{\sqrt{2} L a}{r^6}+\frac{i E}{r^5 \sqrt{E^2-\frac{L^2} {r^2} \left(1-\frac{2 M}{r}\right)}} \right), \nonumber\\
    &\frac{D^2 x^\mu}{D\lambda^2}=0 \qquad \text{for} \qquad \mu \ne 2.\label{te77}
\end{align}
Integrating this equation up to second-order in $M/r$ gives
\begin{equation}
    \dot x^2= \frac{1}{\omega}\left( \frac{C_1}{r}+\frac{2 L M}{E r^4}+ \frac{3 L M^2}{2 E r^5}\right),
\end{equation}
where $C_1$ is an integration constant. The first term, which gives a divergent result upon integration, can be made to vanish by choosing $C_1=0$. We can see that $\dot x^2\ne 0$ for $L\ne 0$, and in that case, the gravitational wave travels along a null nongeodesic trajectory even in the Schwarzschild spacetime, due to the gravitational spin Hall effect.

We can now calculate the angle of deflection of gravitational waves in the $\theta$-direction while passing near a lensing  object of mass $M$, using the relation (here, $x_2= \theta$ and $x_3=\phi$)
\begin{equation}
    d\theta= \frac{\dot\theta}{\dot r} dr= \frac{2 L M}{\omega E r^3 \sqrt{E^2 r^2-L^2}} dr,
\end{equation}
up to leading order in $M/r$. A photon starting from infinity and approaching the lensing object within a closest distance of $R_0$ has $E=1$ and
\begin{equation}
    L= \frac{R_0}{1-\frac{2 M}{R_0}}.
\end{equation}
We can now integrate to obtain the total deflection angle up to leading order in $M/r$,
\begin{equation}
    \theta= 2 \left(\theta_\infty -\theta_0\right)= \frac{\pi M}{\omega R_0^2}.
\end{equation}
This gives an additional deflection resulting from the spin-orbit coupling, which agrees with the result obtained in Ref.~\cite[]{22}, up to a numerical prefactor. For a lensing object of mass $M \ll R_0$, this quantity may be too small to detect by LIGO~\cite[]{36}. This deflection, however, is twice the deflection of electromagnetic waves due to the spin-orbit correction~\cite[]{32}. Moreover, although $\omega$ is only the expansion, it multiplies with the phase ${\cal S}$ in Eq.~\eqref{a11} and thus flipping the sign of $\omega$ from the positive to the negative sign corresponds to the flip in the helicity of waves. As the subleading order corrections are proportional to $\omega$, waves of opposite helicities give the more pronounced spin Hall effect.

\subsection{Propagation of a binary's gravitational waves through an expanding universe}

If we model the large-scale structure of the expanding universe by the line element
\begin{equation}
    ds^2= R_0 t^2 \left(-dt^2+ dr^2+ r^2 d\theta^2+ r^2 \sin^2\theta d\phi^2\right),
\end{equation}
then no spin Hall effect is observed for gravitational waves sourced by a binary system. This is because, in this spacetime, the null rays along which gravitational waves propagate in the geometric optics regime are the curves of constant $\theta$, $\phi$ and $t-r$. A set of null tetrads for such spacetime can be written as
\begin{equation}
    \begin{split}
        & l_{0 \mu}=R_0^2 \left(-1,1,0,0 \right),\\
        & n_{0 \mu}=\frac{t^4}{2} \left(-1,-1,0,0 \right),\\
        & m_{0 \mu}=\frac{R_0 t^2 r}{\sqrt{2}} \left(0,0,1,i \sin \theta \right),\quad \tilde m_{0 \mu}= \left(m_{0 \mu}\right)^*.
    \end{split}
\end{equation}
This is indeed a Fermi propagated null tetrad along the rays that satisfy Eqs.~\eqref{oc29}-\eqref{pt33}. Therefore, these null trajectories $l_0^\mu$ and polarization vectors $m_0^\mu$ constitute the gravitational wave solutions of the linearized Einstein equations in the limit of geometric optics.

Using the propagation Eqs.~\eqref{te56}, we obtain no first-order correction of the null trajectory, thereby implying no spin-orbit coupling for the propagation of gravitational waves in such a spacetime. Thus, the propagation of gravitational waves from a binary does not deviate from the null geodesic trajectories in such spacetime. The absence of terms with a subleading order correction in the geometric optics approximation in such spacetime could be attributed to the fact that waves propagate in the curves of constant $\theta$ and $\phi$. The absence of orbital motion/angular momentum of gravitational waves propagating in such spacetime causes spin-orbit coupling to vanish.

\section{Conclusion and discussions}\label{cd4}

We have presented calculations that account for the spin Hall effect of gravitational waves on curved backgrounds and applied the results to some simple scenarios. The equations of spin optics developed here for gravitational waves are shown to have the exact same form as their electromagnetic counterparts. The only difference is a factor of two, which correctly accounts for the difference in helicity between gravitational and electromagnetic waves. This factor of two causes the deviation of gravitational waves from the null geodesic trajectory to be twice that of electromagnetic waves, due to spin-orbit coupling. It should also be noted that this deviation is frequency-dependent.

In leading-order geometric optics, the trajectory of light/gravitational waves is geodesic. However, this is no longer true when the subleading order correction is considered. Once we include the subleading order correction to the geometric optics, the diffraction effects start to appear and the propagation properties of waves changes. Wave-like effects differ increasingly from the geometric optics as the wavelength increases relative to the length scale of the inhomogeneity. The study of this effect, resulting from the propagation in the Universe with the structure, also helps in disentangling the effects coming from the intrinsic properties of the source or the nature of gravity in the strong field regime. As gravitons/photons still travel at the speed of light, it takes longer for them to reach the observer from the source~\cite[]{32}. Due to their helicity difference, it takes an even longer time for gravitons than for photons. This frequency-dependent time delay for circularly polarized waves might be observed in cosmological events like neutron star mergers~\cite[]{3}. We can also notice from Sec.~\ref{pb} that electromagnetic and gravitational waves in the weak field limit (up to the subleading order geometric optics approximation) follow the same set of equations except for a difference in the factor of two that accounts for their helicity difference. Therefore, the procedure for obtaining the subleading order correction described here can be applied to massless particles of arbitrary spin (obviously, by taking into account the difference in the spin of these particles).

There are different approaches to studying massless particles' motion with spins in curved spacetimes~\cite[]{5}. The similarity of our geometric optics approach to the quantum mechanical approach used by \cite{26} to obtain the gravitational spin Hall effect is evident from the similarity of our Hamiltonian Eq.~\eqref{he65} with theirs. This gives a similar equation for the null trajectory deviating from the geodesic in the subleading order in $1/\omega$, where the term proportional to $b_\beta$ occurring in our Hamiltonian should be interpreted as a Berry connection. Similarly, the equivalence between the quantum mechanical approach, for example, of Gosselin and the approach using Souriau-Saturnini equations~\cite[]{27,28} was demonstrated by Ref.~\cite{5} (at least for the Schwarzschild spacetime). For the Schwarzschild spacetime, the trajectory equations from all of these approaches closely resemble those used in optics to derive the spin Hall effect, where spacetime is treated as an effective medium with perfect impedance matching. This resemblance is particularly encouraging, as the spin Hall effect has been verified experimentally in optics. In future work, we hope to rigorously establish the equivalence of our spin optics approach with these different approaches.


\textbf{Acknowledgements:} We would like to thank Ioannis Soranidis and Phil Simovic for helpful comments. PKD is supported by an International Macquarie University Research Excellence Scholarship.


\appendix

\section{Deriving gravitational wave equations} \label{app0}

For simplification, we first write Eq.~\eqref{fe4} as
\begin{multline}
    \nabla_\mu \nabla^\mu h_{\alpha\beta}- g_{\alpha\beta} \nabla_\mu \nabla^\mu h^\nu_{\nu}+ \nabla_\alpha \nabla_\beta h^\mu_{\mu}\\
    + g_{\alpha\beta} \nabla^\gamma \nabla^\delta h_{\gamma\delta}
    -\nabla^\gamma \nabla_\alpha h_{\gamma\beta}- \nabla^\gamma \nabla_\beta h_{\gamma\alpha}=0.
\end{multline}
Next, we use Eq.~\eqref{tr5} to reduce this equation to
\begin{multline}
    \nabla_\mu \nabla^\mu h_{\alpha\beta}+ \nabla_\alpha \nabla_\beta h^\mu_{\mu}
    -\nabla^\gamma \nabla_\alpha h_{\gamma\beta}\\
    - \nabla^\gamma \nabla_\beta h_{\gamma\alpha}=0. \label{a2}
\end{multline}
Further simplification can be achieved by using Eq.~\eqref{covd9} for the commutation of covariant derivatives
\begin{align}
    \nabla_\gamma \nabla_\alpha h^\gamma_\beta=& \nabla_\alpha \nabla_\gamma h^\gamma_\beta- R^\gamma_{\sigma\alpha\gamma} h^\sigma_\beta+ R^\sigma_{\beta\alpha\gamma} h^\gamma_\sigma\nonumber\\
    =& \nabla_\alpha \nabla_\gamma h^\gamma_\beta+ R^\sigma_{\beta\alpha\gamma} h^\gamma_\sigma,
\end{align}
where we have used the relation $R^\gamma_{\sigma\alpha\gamma}= -R_{\sigma\alpha}= 0$, which follows from Eq.~\eqref{vs2}. Substituting this into Eq.~\eqref{a2} gives
\begin{multline}
    \nabla_\mu \nabla^\mu h_{\alpha\beta}+ \nabla_\alpha \nabla_\beta h^\mu_{\mu}- \nabla_\alpha \nabla_\gamma h^\gamma_\beta- \nabla_\beta \nabla_\gamma h^\gamma_\alpha\\
    - 2 R_{\sigma\beta\alpha\gamma} h^{\gamma\sigma} =0
\end{multline}
Further use of Eq.~\eqref{lc6} gives the linearized gravitational wave Eq.~\eqref{lfe8}.

\section{Reproducing the equations of geometric optics} \label{a31}

From Eqs.~\eqref{lc14}-\eqref{dr16}, we can recover the results of geometric optics by substituting $a_1^{\alpha\beta}=0=l_1^\beta$. In the leading order approximation in $\omega$, the Lorentz condition Eq.~\eqref{lc14} and the wave equation Eq.~\eqref{fe15} gives
\begin{equation}
    l_0^\alpha a_{0 \alpha\beta}= 0= l_0^\alpha l_{0\alpha}.\label{go18}
\end{equation}
This equation allows one to write $a_{0 \alpha\beta}= a_{0\alpha} \left(c_1 l_{0\beta}+c_2 m_{0\beta}\right)$, where $c_1$ and $c_2$ are arbitrary constants, $a_{0\alpha}$ is an arbitrary vector and $m_{0\beta}$ is a complex vector satisfying $m_{0\beta} \tilde m_0^\beta=1$ and $m_{0\beta} l_0^\beta=0$. As the polarization tensor $a_{0 \alpha\beta}$ is symmetric in its indices, we should also have $a_{0\alpha}= c_1 l_{0\alpha}+c_2 m_{0\alpha}$. Now, using the property $\tilde a_0^{\alpha\beta} a_{0 \alpha\beta}=a^2$ we obtain $c_2^2= a e^{i \phi}$. We can further reduce the expression for $a_{0 \alpha\beta}$ by considering the gauge transformation
\begin{equation}
    h'_{\alpha\beta}=h_{\alpha\beta}-\Xi_{\alpha;\beta}-\Xi_{\beta;\alpha},
\end{equation}
under which the linearized Einstein field equations presented above are invariant. Here,
\begin{equation}
    \Xi_\alpha= g_{\alpha\beta} \Xi^\beta= \frac{1}{\omega}\xi_\alpha e^{i {\cal S}}, \label{b3}
\end{equation}
are small arbitrary functions. This gauge transformation preserves the form of WKB ansatz given in Eq.~\eqref{a11} if
\begin{equation}
    a'_{\alpha\beta}=a_{\alpha\beta}-\frac{1}{\omega}\left(\xi_{\alpha;\beta}+\xi_{\beta;\alpha}\right)- i \left(\xi_\beta l_\alpha+ \xi_\alpha l_\beta \right),
\end{equation}
is satisfied. The change in amplitude in the leading order approximation can be written as
\begin{equation}
    a'_{0 \alpha\beta}=a_{0 \alpha\beta}- i \left(\xi_{0 \beta} l_{0 \alpha}+ \xi_{0 \alpha} l_{0 \beta} \right).
\end{equation}
As $l_{0 \alpha} l^{0 \alpha}=0$, the Lorentz condition of Eq.~\eqref{lc14} holds for arbitrary $\xi_{0 \alpha}$ satisfying $\xi_{0 \alpha} l_0^\alpha=0$. This freedom in the choice of $\xi_{0 \alpha}$ can be used to choose $c_1=0$. Moreover, an arbitrary constant parameter $\phi$ from $c_2^2$ can be absorbed in a redefinition of the phase function ${\cal S}$. We finally obtain
\begin{equation}
    a_{0 \alpha\beta}=c_2^2 m_{0\alpha} m_{0\beta}= a m_{0\alpha} m_{0\beta}.\label{pa23}
\end{equation}

Note that it is equally legitimate to write $a_{0 \alpha\beta}= a_{0\alpha} \left(c_2 m_{0\beta}+ c_3 \tilde m_{0\beta}\right)$, where $\tilde m_{0\beta}$ is the complex conjugate of $m_{0\beta}$ and $c_3$ is another arbitrary constant. However, as in electromagnetism, if the polarization tensor with $m_{0\beta}$ represents right-hand circularly polarized waves, then the polarization tensor with $\tilde m_{0\beta}$ represents left-hand circularly polarized waves.

We calculate $\tilde a_{0\alpha\beta}t^{\alpha\beta}$ from Eq.~\eqref{fe15} while considering only the terms relevant to the  geometric optics approximation, that is, we take $a_1^{\alpha\beta}=0$ and $l_{1\mu}=0$ to obtain
\begin{equation}
    l^\mu_{0;\mu}+\frac{2}{a^2} \tilde a_{0 \alpha\beta} a^{\alpha\beta}_{0~~;\mu} l_0^\mu=0.\label{i24}
\end{equation}
Adding this equation to its complex conjugate gives
\begin{equation}
    l^\mu_{0;\mu}+\frac{1}{a^2} \left(\tilde a_{0 \alpha\beta} a_0^{\alpha\beta}\right)_{;\mu} l_0^\mu= \frac{1}{a^2} \left(l_0^\mu {\cal I}_0 \right)_{;\mu}= 0.
\end{equation}
where
\begin{equation}
    {\cal I}_0 \eqdef \tilde a_{0 \alpha\beta} a_0^{\alpha\beta}- \frac{1}{2} \tilde a^\mu_{0\mu} a^\nu_{0\nu}
\end{equation}
is the intensity of the wave (note that $a^\mu_{0\mu} =0$).
These constitute a complete set of equations for gravitational waves in curved spacetime in the limit of geometric optics.

We substitute the value of the polarization amplitude from Eq.~\eqref{pa23} to Eq.~\eqref{i24}, which upon further simplification gives
\begin{equation}
    l^\beta_{0;\beta}+2 \frac{a_{;\beta}} {a} l_0^\beta+ 4 \tilde m_{0\alpha} m^\alpha_{0;\beta}l_0^\beta= 0.
\end{equation}
Since the term $\tilde m_{0 \alpha} m^\alpha_{0;\beta}l_0^\beta$ is purely imaginary and the remaining terms $l^\beta_{0;\beta}+2 l_0^\beta a_{;\beta}/a$ are purely real, they should be separately zero, thereby giving
\begin{equation}
    l^\beta_{0;\beta}+2 \frac{a_{;\beta}} {a} l_0^\beta=0, \qquad  m^\alpha_{0;\beta}l_0^\beta=0. \label{go28}
\end{equation}
These relations will be used below in Appendix~\ref{aa}.

\section{Checking self-duality in the spin optics approximation} \label{aa}

To see that the subleading order solution of the tetrad $\left(\dot x^\alpha, n^\alpha, m^\alpha, \tilde m^\alpha\right)$ given by Eq.~\eqref{so60}, Eqs.~\eqref{so61} and \eqref{so63} is not self-dual, we calculate each of these equations separately, starting with Eq.~\eqref{pe48}
\begin{multline}
    {\cal Z}_{\alpha\beta} l^\alpha \tilde m^\beta=\frac{2 i a}{\omega} \left(-\frac{a_{;\alpha}}{a}l_0^\alpha+m_{0\alpha;\beta}l_0^\alpha \tilde m_0^\beta\right)\\
    =\frac{2 i a}{\omega}\left(\frac{1}{2} l^\alpha_{0;\alpha}-m_0^\alpha l_{0 \alpha;\beta} \tilde m_0^\beta\right) =0,
\end{multline}
where we have used Eq.~\eqref{go28} to arrive at this identity. Similarly, Eq.\eqref{pe47} gives
\begin{multline}
    {\cal Z}_{\alpha\beta} \left(\tilde m^\alpha m^\beta- l^\alpha n^\beta \right)=\frac{2 i a}{\omega} \left(\frac{a_{;\alpha}}{a}m_0^\alpha-m_{0\alpha;\beta}l_0^\alpha n_0^\beta\right)\\
    =\frac{2 i a}{\omega}\left(- m^\alpha_{0; \alpha}- \frac{i}{2} b_\alpha m_0^\alpha-m_{0\alpha;\beta}l_0^\alpha n_0^\beta\right)=0,
\end{multline}
where we have used Eq.~\eqref{lc64} to obtain this identity. Finally, Eq.\eqref{pe46} gives
\begin{multline}
    {\cal Z}_{\alpha\beta} m^\alpha n^\beta=\frac{2 i a}{\omega} \left(-m_{0\alpha;\beta}m_0^\beta n_0^\alpha\right)\\
    =\frac{2 i a} {\omega}\left(m_0^\alpha n_{0\alpha;\beta} m_0^\beta\right) \equiv \frac{2 i a} {\omega} \tilde \lambda,
\end{multline}
where $\tilde \lambda$ is one of the Newman-Penrose scalars. Thus, Eq.\eqref{pe46} is not satisfied unless $\tilde \lambda=0$. Thus, the tetrad $\left(\dot x^\alpha, n^\alpha, m^\alpha, \tilde m^\alpha\right)$, satisfying the scalar products of Eqs.~\eqref{oc29}-\eqref{oc31} does not give the self-dual solution of the field Eqs.~\eqref{fe41} and \eqref{fe42}.

\end{document}